# New Approach to Exact Extraction of Piezoresistance Coefficient

Z. Gniazdowski, B. Latecki and P. Kowalski

Institute of Electron Technology, Al. Lotnikow 32/46, 02-668 Warsaw, Poland
e-mail: gniazd@ite.waw.pl   http://www.ite.waw.pl

***Summary.*** *The extension of piezoresistance coefficient extraction method was proposed, for protection from errors in estimation of the thickness of the test structure membrane. This new approach requires finding models of elements of matrix of integrated stresses as functions of the membrane thickness. Additionally, two test structures with different thickness of the membrane have to be used. This improved method gives the more credible estimation of piezoresistance coefficients $\pi_L$ and $\pi_T$ and the thickness of the membranes of the used test structures.*

## Introduction

The requirements for performances of piezoresistive sensors induce the necessity of optimisation of its functional parameters. Modelling is the significant tool in optimisation process. Longitudinal piezoresistance coefficient $\pi_L$ and transversal piezoresistance coefficient $\pi_T$ are necessary for modelling piezoresistors.

The method for calculating piezoresistance coefficients $\pi_L$ and $\pi_T$ for homogenous layers is known from literature [1], [2], [3], [4]. This method requires well-defined piezoresistance tensor $\Pi$ dependent on the type of conductivity and doping concentration of the semiconductor.

Ion implantation and diffusion is used for fabrication of piezoresistors. In this case modelling with piezoresistance coefficient extracted for uniform doped semiconductor is out of credibility. For that reason, effective longitudinal and transversal piezoresistance coefficients should be extracted for the given technology.

## First approach to extraction problem

The useful approach for solving extraction problem was proposed [5]. For piezoresistor of layout presented in Fig. 1 the resistance $R$ under the stress can be calculated using the following formula [5]:

$$R = R_0 + \rho_0 \pi_L \int_{x_d}^{x_u} \sigma_L(x)dx + \rho_0 \pi_T \int_{x_d}^{x_u} \sigma_T(x)dx \qquad (1)$$

where $\sigma_L(x)$ and $\sigma_T(x)$ are longitudinal and transverse components of the stress along piezoresistor calculated by Finite Element Method (FEM) simulation, $R_0$ is the measured value of piezoresistor without the stress and $\rho_0 = R_0/L$. Denoting:

$$a_L = \int_{x_d}^{x_u} \sigma_L(x)dx, \qquad (2a)$$

$$a_T = \int_{x_d}^{x_u} \sigma_T(x)dx \qquad (2b)$$

also $\Delta R = R - R_0$ and $d = \Delta R/\rho_0$, we get:

$$d = a_L \pi_L + a_T \pi_T. \qquad (3)$$

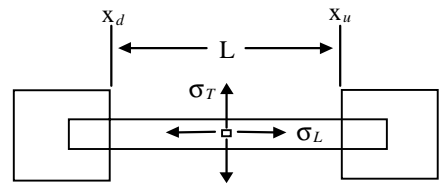

Fig. 1: Layout of the piezoresistor: L is its length, $\sigma_L$ and $\sigma_T$ are stress components.

Considering two completely different resistors on the test structure, we have the system of two linear equations with two unknown piezoresistance coefficients:

$$\begin{bmatrix} d_1 \\ d_2 \end{bmatrix} = \begin{bmatrix} a_{1L} & a_{1T} \\ a_{2L} & a_{2T} \end{bmatrix} \begin{bmatrix} \pi_L \\ \pi_T \end{bmatrix} \qquad (4)$$

In matrix notation, this system has the form:

$$d = A\pi \qquad (5)$$

The resolution of this system for the given technology are values of $\pi_L$ and $\pi_T$.

## Test structure

The adequate test structure is necessary for extraction of piezoresistance coefficients using presented above method. As a test structure, silicon pressure sensor was considered. The structure was fabricated in standard CMOS technology for 5 μm design rule with incorporated anisotropy etching. N-type <100> silicon wafers was applied as a substrate. Four p-type resistors oriented with the [110] crystallographic direction are implanted into the n-type membrane. Piezoresistors are connected into the Wheatstone bridge. The FEM simulation of the sensor was performed for 100-kPa pressure applied to the test structure with the membrane thickness 15μm to 35μm, using SAMCEF system [6]. The exact distributions of the stress components $\sigma_L(x)$ and $\sigma_T(x)$ along piezoresistors are extracted from the simulation results. The distributions of the stress components across the piezoresistors modelled by the second order polynomials are integrated in formula (2), to get the matrix $A$. Elements of this matrix versus test structure membrane thickness are presented in Fig. 2.

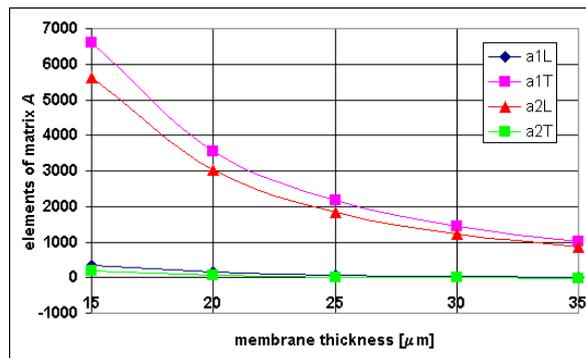

Fig. 2: Elements of matrix A versus membrane thickness

## Conditioning of extraction problem

For the assumed test structure, conditioning of extraction problem was investigated [7]. Assuming no computational error in the solution process, we had the problem: what will be the error in vector $\pi$ for the given errors in matrix $A$ and vector $d$? Estimations well known in numerical analysis were used [8] for solving this problem. For this purpose, the condition number of problem (5) was defined:

$$cond(A) = \| A \| \cdot \| A^{-1} \|, \qquad (8)$$

where $\| x \| = \max_{1 \leq i \leq n} | x_i |$ is the norm of the vector x and

$\| A \| = \max_{1 \leq i \leq n} \sum_{j=1}^{n} | a_{ij} |$ is the corresponding matrix norm

depends on above mentioned vector norm. Condition number is the largest factor by which a relative error in input data can be multiplied when propagate into a relative error in $\pi$. It can be named factor of error propagation. In general, the value of condition number is greater than one. Well-conditioned problems have condition numbers between 1 and 10. They propagate relative errors by a factor no larger than 10. Ill-conditioned problem have condition numbers greater than 100.

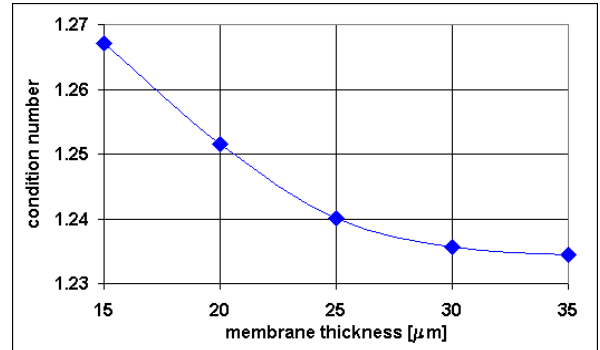

Fig. 3: Condition number versus membrane thickness.

For the given test structure, condition number as a function of its membrane thickness was estimated (Fig. 3). This condition number is less than 1.27. It means, used test structure implies well-conditioned matrix $A$.

## Error propagation

Errors of the extraction with assumptions about some errors of input data were estimated, for the given condition number.

The errors of vector $d$ result from statistical disturbances inherent in technological process. Because of disturbances in the process, the parameters of piezoresistors can change from lot to lot, from wafer to wafer and from chip to chip. Especially, error of $d$ is composed of error of estimation of the length $L$ and errors of resistance $R$ and $R_0$. These errors have a statistical nature. Therefore, they can be considered by manifold extraction of piezoresistance coefficient and statistical estimation.

It becomes to note intuitive obvious fact that the error of $L$ can be partly self-compensated. The changes of $L$ influence the left side of expression (3). In similar way, it influences its right side, by determination of the area of integration in expression (2).

Relative error of matrix $A$ results from error of definition of the thickness of the test structure membrane. FEM simulation is performed for assumed value of the pressure and assumed thickness of the membrane. This thickness can differ from the real thickness resulted from random disturbances in the technological process. Therefore, the matrix $A$ used for solving system (5) can be charged by the error. For example, assuming nominal thickness of the membrane 20 μm and difference between simulated membrane and manufactured membrane about 1 μm, we have the error of matrix A greater than 10%. This error can significantly prevail over the error of vector *d*.

Now the relative error of $\pi$ assuming errors in matrix *A* and vector *d* can be considered. For our consideration, we assume condition number of matrix *A* equal to 1.27, relative error $\|\Delta A\|/\|A\|$ less than 10% and summary relative error $\|\Delta d\|/\|d\|$ less than 1.0%.

If $cond(A) \cdot \|\Delta A\|/\|A\| < 1$, the relationship between their relative errors and relative error of $\pi$ can be estimated by the formula:

$$\frac{\|\Delta \pi\|}{\|\pi\|} \leq \frac{cond(A)}{1 - cond(A) \cdot \|\Delta A\|/\|A\|} \left( \frac{\|\Delta d\|}{\|d\|} + \frac{\|\Delta A\|}{\|A\|} \right). \quad (9)$$

From here, we have total relative error $\|\Delta\pi\|/\|\pi\|$ less than 16%.

If *d* is error free but matrix *A* has error $\Delta A$, we can use the estimation:

$$\frac{\|\Delta \pi\|}{\|\pi\|} \leq \frac{cond(A) \cdot \|\Delta A\|/\|A\|}{1 - cond(A) \cdot \|\Delta A\|/\|A\|}. \quad (10)$$

For that reason, the relative error $\|\Delta\pi\|/\|\pi\|$ is less than 14.55%.

There are two conclusions from this investigation:
1) Proposed method coupled with the given test structure is well-conditioned. It means that input data error propagation factor is near one (Fig. 3).
2) If the input data is charged by an error then the result of extraction is charged by at least the same error like the input data error. This is true also for well-conditioned method of extraction.

### Improvement of the method

Above mentioned analysis shows that considered method has some disadvantage. The results of piezoresistance coefficient extraction can be charged by some significant error. It follows from the error in the estimation of the thickness of the test structure membrane. We propose the improvement of the presented above extraction method, for protection from this disadvantage.

Piezoresistance coefficient is a function of the property of the piezoresistive layer. On the other hand, it is not dependent on the thickness of the membrane. It means: for two test structures with different membrane thickness manufactured in the same technological process, the results of extraction have to be the same.

For assumed nominal value of the pressure the elements of matrix *A* are non-linear functions of the membrane thickness *t* (Fig. 2). These dependencies can be modelled for the assumed range of *t* changes, using regression analysis method.

Proposed algorithm consists of several steps:

1. Calculate components of matrix *A*, for several values of *t* located in assumed range of changes. Model these components as dependent on *t* nonlinear functions, for instance polynomials.
2. Select two test structures with explicitly different thickness of the membrane.
3. Measure two resistors on both test structures without the pressure and with the nominal value of the pressure. Evaluate vectors *d* for both test structures.
4. Scan the entire domain of changes of the thickness *t* for both membranes. For all couples $(t_1,t_2)$ estimate the pair of matrixes *A* using models fitted in step 1. Resolve the system (5) for both matrixes. Estimate the distance between obtained vectors $\pi$, in a sense of Euclidean norm.
5. The couple $(t_1,t_2)$ for which the calculated distance is the smallest is the estimation of the membrane thickness of the used test structures. As a result of extraction, the average vector of both vectors $\pi_1$ and $\pi_2$ obtained for this couple can be received.

Presented above algorithm was used for extraction of piezoresistance coefficient. The results of measurement of piezoresistors and results of extraction are presented in Table 1.

Table 1.

|  |  | *Membrane thinner* | *membrane thicker* |
|---|---|---|---|
| $R_1[\Omega]$ | 0kPa | 4724.88 | 4795.64 |
|  | 100kPa | 4492.96 | 4675.32 |
| $R_2[\Omega]$ | 0kPa | 4740.80 | 4728.00 |
|  | 100kPa | 4979.80 | 4851.20 |
| *Results* | | | |
| *t* [μm] | | 16.450 | 22.620 |
| $\pi_L [10^{-4} MPa^{-1}]$ | | 6.947 | 7.057 |
| $\pi_T [10^{-4} MPa^{-1}]$ | | -5.967 | -5.967 |
| *Average* | | $\pi_L = 7.00 \, [10^{-4} MPa^{-1}]$ $\pi_T = -5.97 \, [10^{-4} MPa^{-1}]$ | |

### Conclusions

The extension of the piezoresistance coefficient extraction method was proposed. This new approach can be used for protection from errors in estimation of the thickness of the test structure membrane. This method requires to find nonlinear models of elements of a certain matrix *A* as a functions of the membrane thickness and use two test structures with different thickness of the membrane. The estimation of piezoresistance coefficients $\pi_L$ and $\pi_T$ and the thickness of the used test structures membranes are obtained as a result of extraction.

**T2P28**
Mechanical and thermal sensors

*Z. Gniazdowski et al., New Approach to Exact Extraction
of Piezoresistance Coefficient, pp. 523-526***Acknowledgement**

The authors would like to express their gratitude to Mr Jan Koszur and Dr Jerzy Weremczuk for theirs constructive suggestions. This work was supported by the project No. 8T11B03116 from the State Committee for Scientific Research (KBN).**References**

[1] W. Gopel, J. Hesse, J.N. Zemel, "Sensors. A Comprehensive Survey". Vol. 7. "Mechanical Sensors". Edited by H.H. Bau, N.F. de Rooij, B. Kloeck, VCH, Weinheim 1994

[2] O.N. Tufte, D. Long, "Recent Developments in Semiconductor Piezoresistive Devices", Solid-State Electronics, Pergamon Press 1963. Vol. 6, pp. 323-338

[3] Y. Kanda, "Piezoresistance effect of silicon", Sensors and Actuators A, Vol. 28 (1991) 83-91

[4] Y. Kanda, "A Graphical Representation of the Piezoresistance Coefficients in Silicon", IEEE Trans. on Electron Devices, Vol. ED-29, No. 1, Jan. 1982

[5] Z. Gniazdowski, P. Kowalski. Practical approach to extraction of piezoresistance coefficient. Sensors and Actuators A, Vol. 68 (1998) 329-332

[6] Z. Gniazdowski, J. Koszur and P. Kowalski. Conditioning of piezoresistance coefficient extraction. MIXDES 2000. 7th International Conference Mixed Design of Integrated Circuits and Systems, Gdynia, Poland, 15-17 June 2000

[7] SAMCEF Users Manuals, Samtech, Liege 1996 and 1999

[8] S.M. Pizer, "Numerical computing and Mathematical Analysis". Science Research Associates, Inc. 1975**526**